\documentclass{PoS}

\title{Seasonal variation of atmospheric muons in IceCube}

\ShortTitle{Muon seasonal variations}

\author{
The IceCube Collaboration\footnote{For collaboration list, see PoS(ICRC2019) 1177.}\\
{\itshape \href{http://icecube.wisc.edu/collaboration/authors/icrc19_icecube}{http://icecube.wisc.edu/collaboration/authors/icrc19\_icecube}}\\
E-mail: \email{tilav@udel.edu, gaisser@udel.edu}
}

\abstract{
After more than seven years of data taking with the full IceCube detector 
triggering at an average rate of 2.15 kHz, a sample of
half a trillion muon events is available for analysis.  The extreme
temperature variations in the stratosphere together with the high data rate
reveal features on both long and short time scales with unprecedented precision.
In this paper we report an analysis in terms of the atmospheric profile for production
of muons from decay of charged pions and kaons.  We comment on the 
implications for seasonal variations of neutrinos, which are presented in
a separate paper at this conference.\\

\vspace{4mm}
{\bfseries Corresponding authors:}
\speaker{Serap Tilav}$^{1}$, Thomas K. Gaisser$^{1}$, Dennis Soldin$^{1}$, Paolo Desiati$^{2}$\\
{$^{1}$ \itshape Bartol Research Institute, Dept. of Physics and Astronomy, University of Delaware, Newark, DE 19716 USA}\\
{$^{2}$ \itshape Wisconsin Institute for Particle Astrophysics and Cosmology, University of Wisconsin, Madison, WI 53706 USA}
}

\FullConference{36th International Cosmic Ray Conference -ICRC2019-\\
		July 24th - August 1st, 2019\\
		Madison, WI, U.S.A.}

\begin{document}

\section{Introduction}\label{sec:intro}

Measurements of seasonal variations of the muon flux in deep underground detectors have
a long history, starting with a detector in a deep cavity near Cornell
 University~\cite{Barrett:1952woo}.  The papers reporting the results from the MINOS
 far~\cite{Adamson:2009zf} and near~\cite{Adamson:2014xga} detectors review data from several experiments in terms of minimum
 muon energy needed to reach the detector (e.g. $0.73$~TeV at the MINOS far detector at Soudan
 and $\sim 50/\cos{\theta}$~GeV for the near detector at Fermilab).  The variations are
 characterized by a correlation coefficient $\alpha_T(E_\mu)$ obtained by fitting a straight line
 to rate vs. effective temperature at the energy of each detector.  The correlation coefficient
 and the effective temperature are defined respectively as
 \begin{equation}
 \frac{\delta R}{\langle R\rangle} = \alpha_T\times \frac{\delta T}{\langle T\rangle}
\label{eq:alpha_T}
 \end{equation}
 and
\begin{equation}
T_{\rm eff}(\theta) = \frac{\int {\rm d}E_{\mu}\,\int dX\,P_{\mu}(E_{\mu}, \theta, X)\,A_{\rm eff}(E_{\mu}, \theta)\,T(X)}{\int {\rm d}E_{\mu}\,\int dX\,P_{\mu}(E_{\mu}, \theta, X)\,A_{\rm eff}(E_{\mu}, \theta)}.
\label{eq:teff}
\end{equation}

For compact underground tracking detectors, $A_{\rm eff}$ is simply the projected fiducial area
of the detector coupled with the selection efficiency of single tracks.  For IceCube, with its widely
spaced detectors the effective area requires a Monte Carlo simulation of the detector response
to the class of events used for the analysis.  
A significant technical difference is that the MINOS analysis is done in terms of integral
quantities that refer to all muons above a minimum energy, while the IceCube analysis is differential
in muon energy.  Thus for IceCube the muon production spectrum $P_{\mu}(E_{\mu}, \theta, X)$ 
in Eq.~\ref{eq:teff} is number of muons produced 
per logarithmic bin of energy per g/cm$^2$ of slant depth $X$ along a trajectory
at zenith angle $\theta$.

The basic physics responsible for the seasonal variation of the muon flux and its 
dependence on energy in the region $\sim 50$~GeV to $5$~TeV is the competition between
interaction and decay for the charged pions and kaons that are the dominant source of
muons (and muon neutrinos) in this energy region.  As temperature increases, the atmosphere
expands and decay to muons becomes more likely compared to re-interaction of the parent meson.
The critical energy of a hadron is 
the energy at which decay and interaction have equal probability at a slant depth
comparable to the interaction length.  The relation between density and atmospheric depth (pressure)
depends on temperature through the ideal gas law, leading to the expression for
the critical energy parameter as a function of temperature at depth $X_{Vertical}=X/cos(\theta)$:
\begin{equation}
\epsilon_i(X) = \frac{R\,T(X)}{Mg}\frac{m_ic^2}{c\tau_i},
\label{eq:Ecritical}
\end{equation}
where $M=0.028964$~kg mol$^{-1}$ for dry air, $g$ is the acceleration of gravity and $R=8.3144$ J K$^{-1}$mol$^{-1}$. 
For $T=220^\circ$~K, $\epsilon_\pi = 115$~GeV and $\epsilon_K=857$~GeV.

Although there is some uncertainty in relating the theoretical formalism for inclusive
fluxes of single muons that we use to the IceCube data sample described in the next section,
we demonstrate in this paper that the high event rate of 190 million events per day allows
unprecedented resolution of features in the muon flux.  The large size of IceCube also 
makes possible a measurement of seasonal variations of $\nu_\mu$~\cite{Gaisser:1412971,Zoecklein:aachen}.  
In the concluding
section of this paper, we comment on the complementarity provided by these two 
measurements, in particular in connection with the ratio of kaons to pions in the secondary
cosmic radiation.
 \begin{figure}
\includegraphics[width=0.99\textwidth]{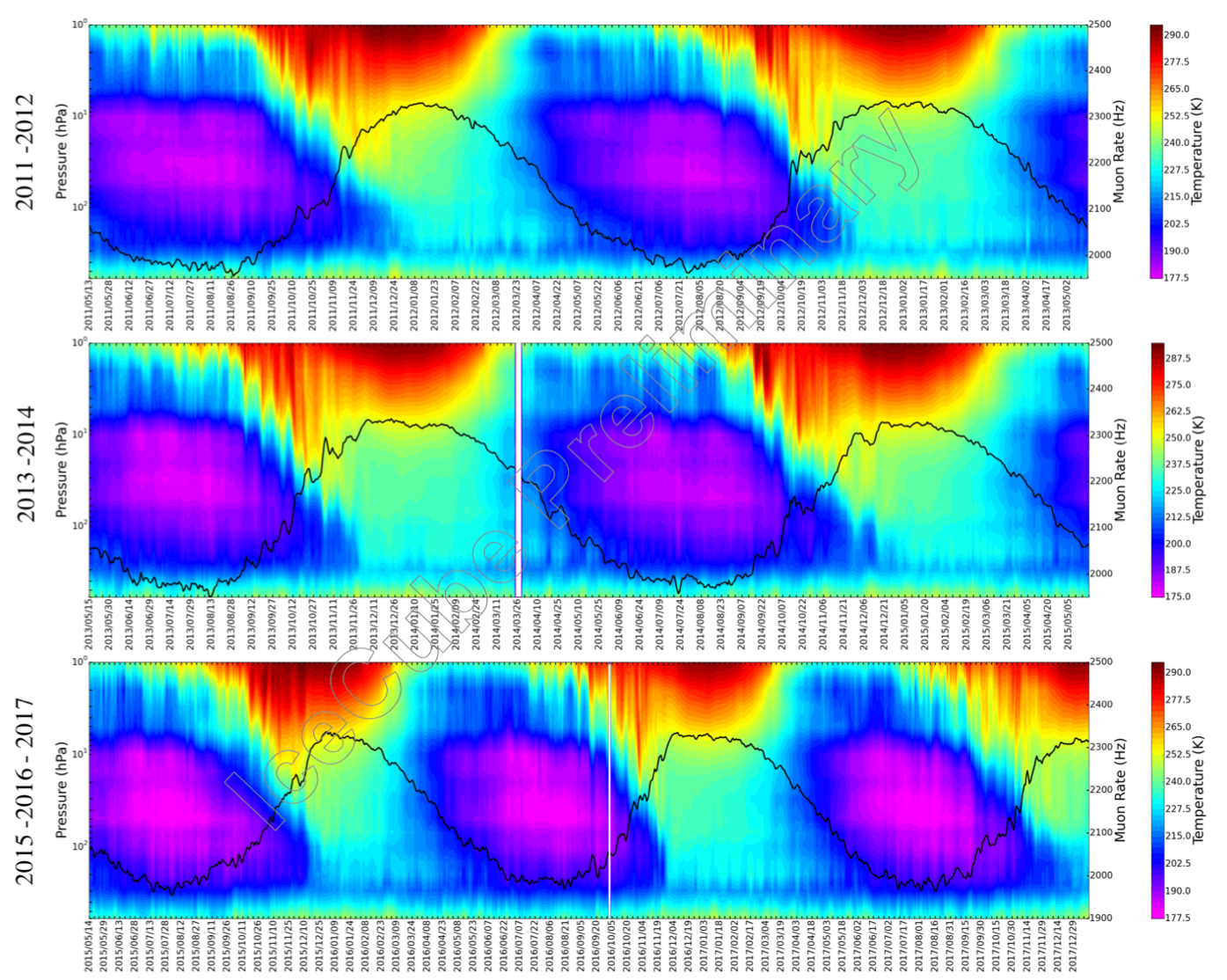}
\caption{IceCube Muon Rate (black line) overlaid with the temperature profile of the South Pole atmosphere at different pressure heights. The plot illustrates the behavior of the seasonal cycles as well as the short-term (day to week time scales) variations in rate with respect to the temperature variations in the stratosphere.}
\label{fig:7yearswithtemps}
\end{figure}
\section{Measurement of muons in IceCube}
The IceCube Neutrino Observatory, located at the geographical South Pole, records high-energy muons at depths of 1450-2450m in the Antarctic ice. While the sensors look for rare astrophysical neutrinos as signal, downgoing muons with energies above 400 GeV are able to penetrate and trigger the detector at a rate of 2.15 kHz on average with $\pm 8\%$ seasonal variation.

Events that pass the InIce-SMT8 trigger criterion in IceCube (a simple multiplicity trigger of 8 or more sensors with local coincidence in 5 $\mu$sec) are used for this analysis.  A previous
discussion of the analysis based on data from IceCube during construction from
2007 to 2011 (IC22, IC40, IC59 and IC79) was presented in~\cite{Desiati:2011hea}.  Here we use seven years of data taken with the fully completed detector since May 2011 (IC86). The daily rate is obtained by using only the runs longer than 30 minutes with complete detector configuration. Fig.~\ref{fig:7yearswithtemps} illustrates the IceCube muon rate correlation with the temperature profile of the South Pole atmosphere over 7 years. The South Pole atmospheric temperature profile is extracted from data supplied by the AIRS~\cite{AIRS} (Atmospheric Infra Red Sounder) instrument aboard NASA's Aqua satellite. 

\section{Muon production profile and effective temperature} 
\label{sec:Teff}
The effective temperature for each day is obtained by weighting the temperature profile in the atmosphere
with the muon production spectrum along the trajectory at zenith angle $\theta$ and integrating over angle.
Low- and high-energy forms for the production spectrum are respectively~\cite{Gaisser:2016uoy}
\begin{equation}
\label{mu-prod-low}
P_{\mu}(E_\mu,\theta,X)\;\approx\;N_0(E_\mu)\,{e^{-X/\Lambda_N}\over \lambda_N}\times
\left[ Z_{N\pi}\,Z_{\pi\mu}(\gamma) +0.635\,Z_{NK}Z_{K\mu}(\gamma)\right ]
\end{equation}
and
\begin{eqnarray}
P_{\mu}(E_\mu,\theta,X)\,\approx&N_0(E_\mu)& \left\{ {\epsilon_\pi \over X\cos\theta\,E_\mu}\right.\, 
Z_{\pi\mu}(\gamma+1) \,
{Z_{N\pi}\over 1-Z_{NN}}\,{\Lambda_\pi\over \Lambda_\pi - \Lambda_N} \nonumber \\
&&\times
\left(e^{-X/\Lambda_\pi}\,-\,e^{-X/\Lambda_N}\right) \nonumber \\
&+0.635&{\epsilon_K \over X\cos\theta\,E_\mu}\,
Z_{K\mu}(\gamma+1) \,
{Z_{NK}\over 1-Z_{NN}}\,{\Lambda_K\over \Lambda_K - \Lambda_N} \nonumber \\
&&\times
\left(e^{-X/\Lambda_K}\,-\,\left. e^{-X/\Lambda_N}\right)\right\}.
\label{mu-prod-high}
\end{eqnarray}
Each equation has one term for muons from decay of charged pions (branching ratio $\approx 1$)
and another for charged kaons (branching ratio $\approx 0.635)$. The nucleon interaction and attenuation lengths are related as  $\lambda_N = \Lambda_N (1-Z_{NN})$. 
The spectrum weighted moments
have the form $Z_{ab} = \int\,x^\gamma\frac{{\rm d}n_{ab}}{{\rm d}x}{\rm d}x$ for $a\rightarrow b$.  
The moments for production of pions and kaons depend on the model of hadronic interactions
used to describe production of pions and kaons by interactions of cosmic-ray nucleons
in the atmosphere, while the decay moments depend only on the two-body decay kinematics
of pion and kaon decay.  In particular,
\begin{equation}
Z_{\pi \mu}(\gamma)\,=\,\frac{(1-r_\pi^{\gamma+1})}{(\gamma+1)(1-r_\pi)}\,=\,\int_{r_\pi}^1 x^\gamma\frac{{\rm d}n_\mu}{{\rm d}x}{\rm d}x
\label{eq:pi2mu}
\end{equation}
and
\begin{equation}
Z_{\pi\mu}(\gamma+1)\,=\,\frac{(1-r_\pi^{\gamma+2})}{(\gamma+2)(1-r_\pi)},
\label{pi2mu-2}
\end{equation}
where $x = E_\mu/E_\pi$, $\gamma$ is the integral spectral index of the cosmic-ray spectrum and $r_\pi = (m_\mu/m_\pi)^2\approx 0.573$.
The forms for two-body decay of charged kaons are the same but with $r_K=(m_\mu/m_K)^2\approx 0.046$. For the calculations of this study we use the Sibyll 2.3c hadronic interaction model~\cite{Riehn:2017mfm} and the H3a model~\cite{Gaisser:2011cc} for nucleon fluxes.

The forms \ref{mu-prod-low} and \ref{mu-prod-high} are combined in the approximation
\begin{equation}
P_\mu(E_\mu,\theta,X)\,=\,\frac{P_{\pi,\rm low}}{1+P_{\pi,\rm low}/P_{\pi,\rm high}}\,
+\,\frac{P_{K,\rm low}}{1+P_{K,\rm low}/ P_{K,\rm high}}
\label{mu-prod}
\end{equation} 
and integrated using Eq.~\ref{eq:teff} to obtain the effective temperature for each direction.
The denominator of Eq.~\ref{eq:teff} is the rate of events, which normalizes the effective temperature.
Finally, the weighted sum over zenith angle gives the effective temperature.
The dependence on temperature comes entirely from the temperature dependence of
the critical energies shown in Eq.~\ref{eq:Ecritical}.

\section{Correlation with effective temperature}
\label{sec:correlation}

Temperature profiles at the South Pole are obtained from the AIRS satellite system
at 21 atmospheric depths from 1 to 800 hecto-pascals in quasi logarithmic intervals.
We use these temperature profiles to calculate event rate and $T_{\rm eff}$ for each day.
\begin{figure}
\centering
\includegraphics[width=0.95\textwidth]{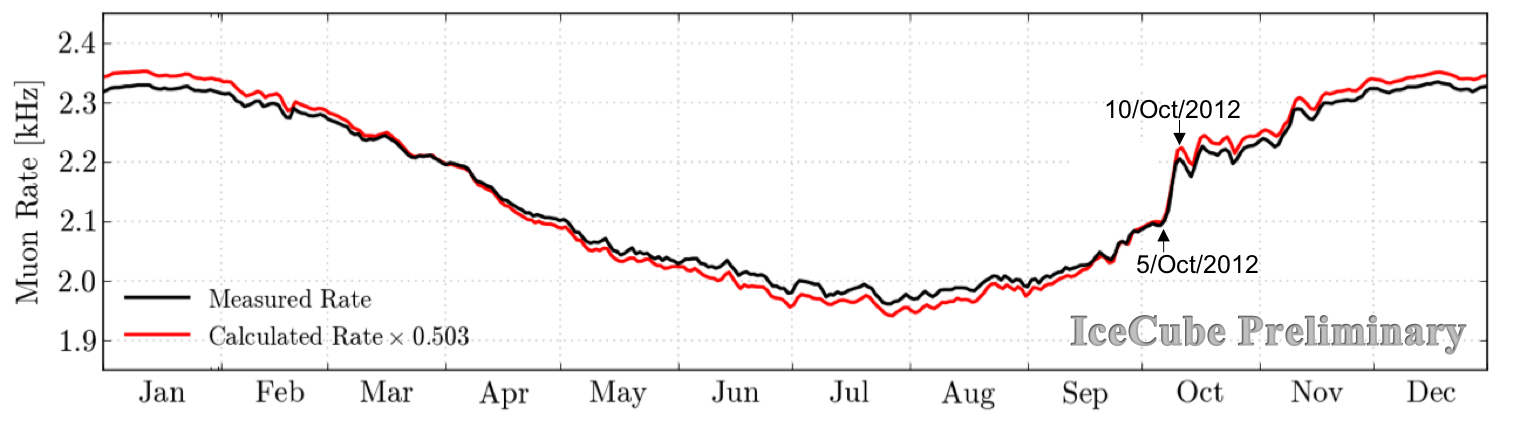}
\caption{Comparison of measured muon rate with calculated rate for 2012.}
\label{fig:data_calc}
\end{figure}
As an example, we show in Fig.~\ref{fig:data_calc} the comparison of the measured and the calculated rate for 2012. The calculated rate depends on the primary spectrum of nucleons evaluated at the energy of the muon ($N_0(E_\mu)$) and on $A_{\rm eff}$ and is normalized here to the observed rate. The calculation matches the features well, but
is off by a factor of two in absolute rate.  The normalization is directly proportional to the primary spectrum of nucleons, so a revised calculation starting with direct measurement to normalize the primary spectrum is underway.  The calculated amplitude of the seasonal variation (maximum rate divided by minimum rate) is $\approx 2\%$ greater in the calculation than measured, but the
 short-term features agree remarkably well. The observed sudden rate jump by 5.4\% in 5 days during 5-10/Oct/2012 is reproduced in the calculated rate as 5.9\% increase, which is caused by the 7.2\% increase in $T_{\rm eff}$ during the same days. Sudden rate jumps of this magnitude are not uncommon during the early October period of each year, as seen in Fig.~\ref{fig:7yearswithtemps}, although the increase in 2012 is exceptionally sharp.  

\begin{figure}
\centering
\includegraphics[width=0.99\textwidth]{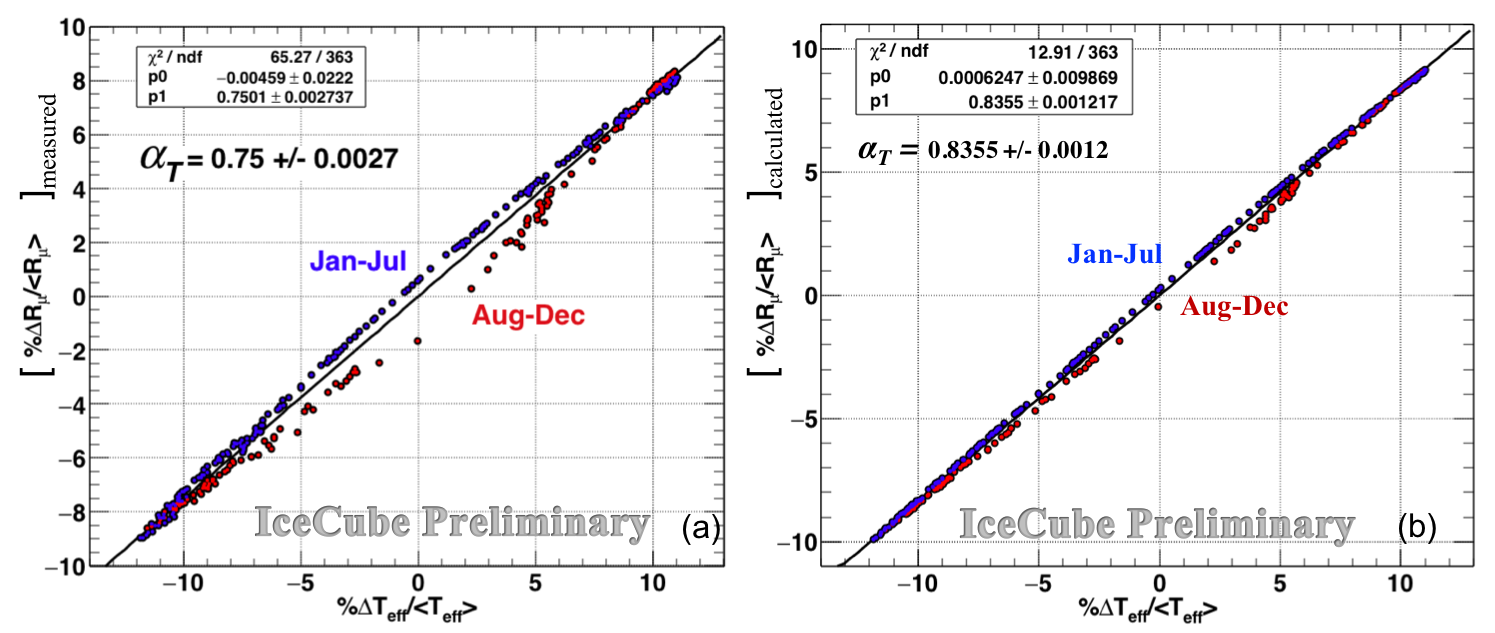}
\caption{Correlation coefficient between (a): the measured rate and $T_{\rm eff}$ (b): the calculated rate and $T_{\rm eff}$ for InIce-SMT8 muon events of IceCube 2012 data. }
\label{fig:2012data}
\end{figure}

Fig.~\ref{fig:2012data}(a) gives the correlation between the measured rate and the calculated $T_{\rm eff}$ for 2012 showing the \% variation of the measured muon rate $R_{\mu}$ over the average rate vs the \% variation of $T_{\rm eff}$ with respect to $\langle T_{\rm eff}\rangle$ for 2012.
From Eq.~\ref{eq:alpha_T},
the slope indicated by the line in the figure is the correlation coefficient $\alpha_T$.
The experimental value of the correlation coefficient is 0.75, which is
about what is expected for the $\sim$TeV muons that dominate the InIce-SMT8 trigger.  To illustrate the non-linearity of the relation between muon rate and effective temperature
we show in Fig.~\ref{fig:2012data}(b) the calculated $\delta R_\mu/\langle R_\mu\rangle$ vs the same $\delta T_{\rm eff}/\langle T_{\rm eff}\rangle$ 
used in Fig.~\ref{fig:2012data}(a) for the measured rate.  The calculated rate shows a qualitatively similar, though slightly smaller, hysteresis than the measured rate. The observed hysteresis exhibits a characteristic behavior for the South Pole related to the qualitatively different temperature profile in the Austral Spring.  The upper atmosphere warms quickly while deeper in the atmosphere the air remains cold. 
The calculated slope corresponds to
an $\alpha_T\approx 0.84$.  Its larger value corresponds to the slightly larger annual modulation
of the calculated rate in Fig.~\ref{fig:data_calc}.
What is new here is that the high precision of the IceCube rates with statistical
fluctuations at the level of $10^{-4}\times \langle R_\mu\rangle$, makes visible for the first time
the non-linearity of the relation between rate and effective temperature.  

To illustrate the origin of this non-linearity it is helpful to go through the analysis explicitly at fixed muon energy where $A_{\rm eff}$ and primary flux cancel.  We use an energy of 1 TeV, which is characteristic of muons in IceCube at 2 km depth in ice.  For fixed energy the muon flux at zenith angle $\theta$ and atmospheric depth $X_0$ is
\begin{equation}
    \phi_\mu(E_\mu,\theta)\,=\,N_0(E_\mu)
\int_0^{X_0/\cos\theta}\left\{\frac{A_{\pi\mu}(X)}{1+B_{\pi\mu}(X)E_\mu\cos\theta/\epsilon_\pi}+\frac{A_{K\mu}(X)}{1+B_{K\mu}(X)E_\mu\cos\theta/\epsilon_K}\right\}{\rm d}X,
\label{eq:muflux}
\end{equation}
where
\begin{equation}
    A_{M\mu} = R_{M\mu}Z_{NM}Z_{M\mu}(\gamma)\frac{\exp{(-X/\Lambda_N)}}{\lambda_N}
    \label{eq:A-factor}
\end{equation}
and
\begin{equation}
 B_{M\mu}=\frac{Z_{M\mu}(\gamma)}{Z_{M\mu}(\gamma+1)}\frac{\Lambda_M-\Lambda_N}{\Lambda_M\Lambda_N}\frac{Xe^{-X/\Lambda_N}}{e^{-X/\Lambda_M}-e^{-X/\Lambda_N}}.
\label{eq:B-factor}
\end{equation}
Here $R_{M\mu}$ is the branching ratio of meson $M = \pi$ or $K$ to muons.  The dependence on temperature is contained entirely in the critical energies in Eq.~\ref{eq:muflux} as defined in Eq.~\ref{eq:Ecritical}.
From its definition in Eq.~\ref{eq:alpha_T}, the correlation coefficient can be calculated from the derivative with respect to $T$ of the rate $R$ as
\begin{equation}
   \alpha_T(E,\theta)\,=\,\frac{\langle T\rangle}{\langle R\rangle} \frac{{\rm d}R}{{\rm d}T}.
   \label{eq:alpha_theo}
\end{equation}
The rate $R$ is proportional to Eq.~\ref{eq:muflux}, so
\begin{equation}
    \frac{{\rm d}R}{{\rm d}T}=N_0(E_\mu)\int\left\{\frac{A_{\pi\mu}B_{\pi\mu}E_\mu\cos\theta/\epsilon_\pi(\langle T\rangle)}{(1+B_{\pi\mu}E_\mu\cos\theta/\epsilon_\pi)^2}+\frac{A_{K\mu}B_{K\mu}E_\mu\cos\theta/\epsilon_K(\langle T\rangle)}{(1+B_{K\mu}E_\mu\cos\theta/\epsilon_K)^2}\right\}\frac{\langle T\rangle}{T^2(X)}{\rm d}X.
\end{equation}
Multiplying by $\langle T\rangle/\langle R\rangle$, the flux $N_0(E_\mu)$ cancels, and $\alpha_T(E,\theta)$ follows from Eq.~\ref{eq:alpha_theo}.

\begin{figure}
\includegraphics[width=0.99\textwidth]{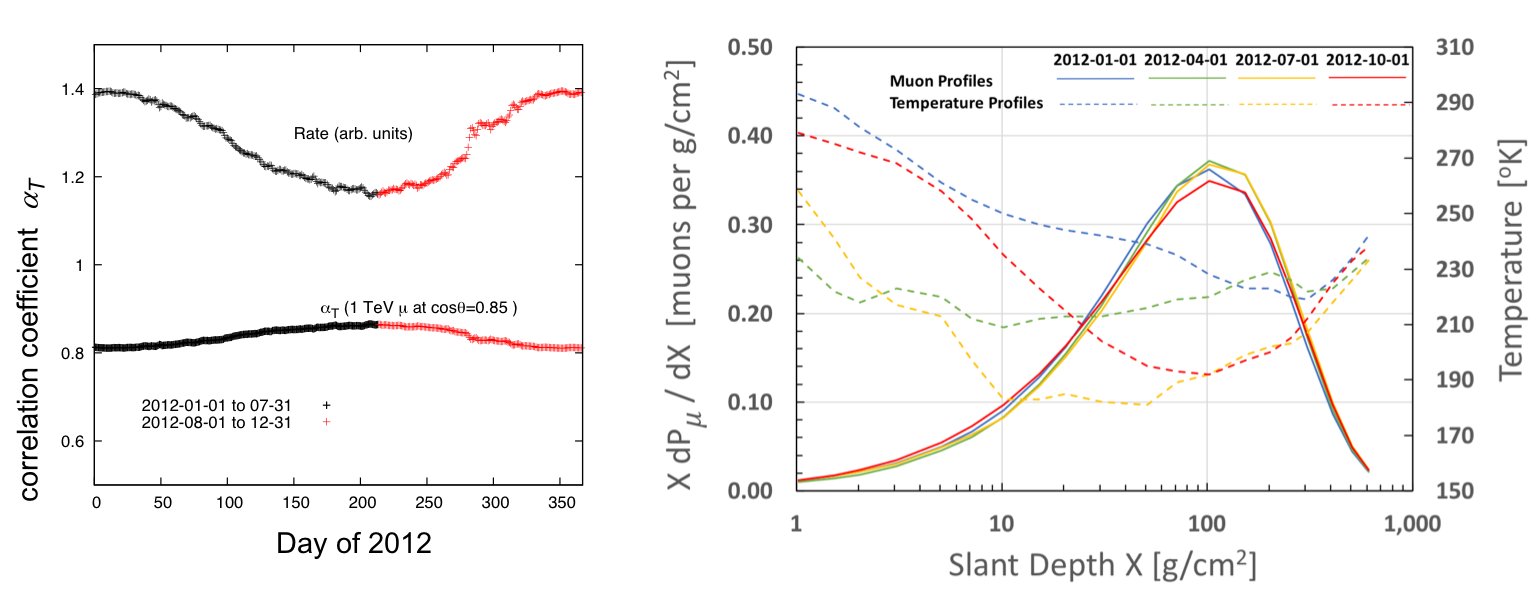}
\caption{Left: Rate and correlation coefficient for TeV muons at $\cos\theta=0.85$. Right: muon production and temperature profiles for 4 days at the beginning of each quarter of 2012.}
\label{fig:profiles}
\end{figure}

The result of the calculation is shown in the Fig.~\ref{fig:profiles}.  The correlation coefficient in the left plot is the slope of Rate vs.
$T_{\rm eff}$ and should be compared with Fig.~\ref{fig:2012data}.  The correlation coefficient for each day is the convolution of the muon production profile and the temperature profile.  The four days shown in the right panel of Fig.~\ref{fig:profiles} are chosen to illustrate seasonal differences.  In particular, the October temperature profile with its high value in the upper atmosphere leads to a lower rate for the same $T_{\rm eff}$ compared to April.


\section{Discussion}
\label{sec:conclusion}
%
%
%
%

Production spectra for $\nu_\mu +\overline{\nu}_\mu$ have the same form as for muons.  The only difference 
comes from the decay factors for the parent mesons.  For the neutrino from meson $M$ ($\pi^\pm$ or $K^\pm$),
\begin{equation}
    Z_{M\nu}(\gamma) = \frac{(1-r_M)^{\gamma+1}}{(\gamma+1)(1-r_M)}
    \label{eq:M1}
\end{equation}
and
\begin{equation}
Z_{M\nu}(\gamma+1)\,=\,\frac{(1-r_M)^{\gamma+2}}{(\gamma+2)(1-r_M)}.
\label{eq:M2}
\end{equation}
Because $r_\pi$ is large, the muon carries most of the energy in pion decay, while in kaon decay the energy is shared almost equally between the muon and the neutrino.  As a consequence, the kaon channel becomes the dominant source of $\nu_\mu$ above $\sim 100$~GeV where Eq.~\ref{eq:M2} applies.  This feature means that the study of
seasonal variations of muons and neutrinos in the same framework should be most sensitive to features like the
kaon to pion ratio.

\begin{figure}
\centering
\includegraphics[width=0.6\textwidth]{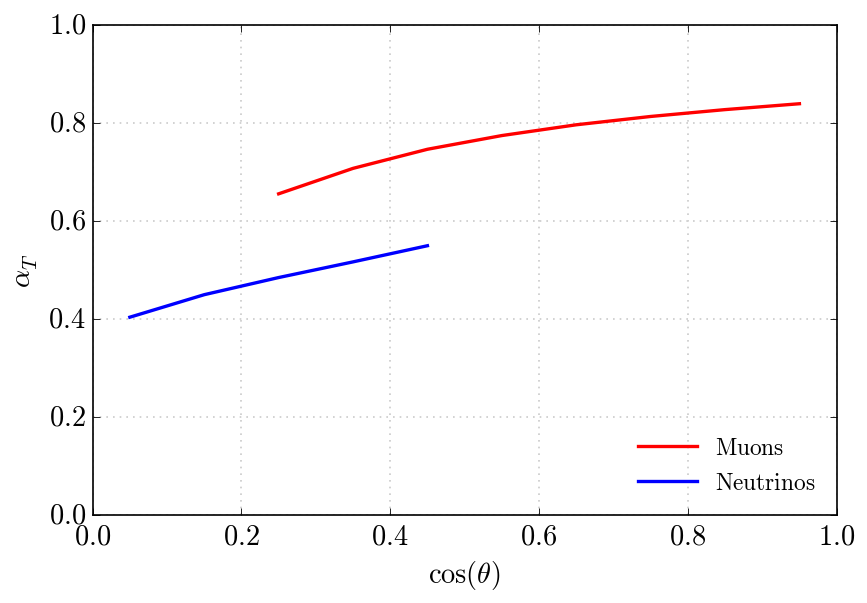}
\label{fig:tempcoeff}
\caption{Zenith angle dependence of the calculated correlation coefficient. }
\end{figure}

To illustrate this possibility we calculate the correlation coefficient for muons and neutrinos at 1~TeV,
an energy that is typical for data samples for seasonal variations of neutrinos as well as for muons in IceCube.
Analysis of seasonal variations of neutrinos in IceCube~\cite{Gaisser:1412971,Zoecklein:aachen} is done with  a data sample of upgoing neutrino-induced muons from the Southern sky with
zenith angles 90$^\circ$ to 120$^\circ$.  The temperatures relevant for the neutrinos cover a much larger portion
of the sky than for the downward muons at the South Pole, for which the zenith angle range is $\approx 0^\circ$ to $60^\circ$.  For simplicity therefore we estimate the correlation coefficients by making the calculation at  fixed $T=220^\circ$~K.
  The angular dependence of the correlation coefficients at $1$~TeV are compared for muon neutrinos and for muons in Fig.~\ref{fig:tempcoeff}.  The atmospheric neutrino flux is largest near the horizon.  The important region for downward muons is near the vertical 
$\cos\theta\ge 0.5$.  

{\bf Summary:} The large volume of IceCube allows study of seasonal variations of neutrinos as well as muons.  The high rate of muons in IceCube provides a statistical precision of the data that reveals significant variations on short time scales,
as illustrated in Fig.~\ref{fig:data_calc}.   The high precision also reveals the non-linearity in the relation between
rate and effective temperature illustrated in Fig.~\ref{fig:2012data}.  Work in progress includes updating the primary spectrum and 
revisiting the calculation of
effective area to account for details of the data selection, for accidental coincident events and for the small contribution of multiple muons to the signal.

\bibliographystyle{ICRC}
\bibliography{seasonal}

\end{document}